\documentclass[structabstract]{aa}  
\usepackage{graphicx}
\usepackage{txfonts}

\usepackage{natbib}
\bibpunct{(}{)}{;}{a}{}{,} 

\newcommand{\shortitle}{Integrated spectra extraction based on S/N optimization using IFS}

\usepackage{amssymb}
\usepackage[ps2pdf,bookmarksopen,breaklinks=true]{hyperref}

\hypersetup{citecolor={blue},linkcolor={blue},urlcolor={blue}}     

\hypersetup{
  colorlinks=true,             
  bookmarksnumbered=true,      
  bookmarksopen=false,         
  pdfpagemode=UseOutlines,     
  pdfdisplaydoctitle,          
  pdftitle={\shortitle}
}

\newcommand{\nii}{[N\,{\footnotesize II}]}

\newcommand{\ha}{H$\alpha$}

\defcitealias{Arribas:2008p3550}{A08}

\begin{document}
%

   \title{Integrated spectra extraction based on signal-to-noise optimization using Integral Field Spectroscopy}


   \author{F. F. Rosales-Ortega,
          \inst{1,2}
          S. Arribas
          \inst{1}
          \and
          L. Colina
          \inst{1}
          }

   \institute{Centro de Astrobiolog{\'i}a (CSIC-INTA),
     Ctra de Ajalvir km 4, 28850, Torrej{\'o}n de Ardoz, Madrid, Spain.
     \and
     Departamento de F{\'i}sica Te{\'o}rica, Universidad Aut\'onoma de Madrid,
     28049 Madrid, Spain.\\
     \email{frosales@cantab.net, arribas, colina@cab.inta-csic.es}
             }

   \date{Received July, 2011; accepted December, 2011}

 
  \abstract
   {}
   {We propose and explore the potential of a method to extract high
     signal-to-noise (S/N) integrated spectra related to physical and/or
     morphological regions on a 2-dimensional field using Integral Field
     Spectroscopy (IFS) observations by employing an optimization procedure
     based on either continuum (stellar) or line (nebular) emission features.}
   {The optimization method is applied to a set of IFS VLT-VIMOS observations of
     (U)LIRG galaxies, describing the advantages of the optimization by
     comparing the results with a fixed-aperture, single spectrum case, and by
     implementing some statistical tests.}
   {We demonstrate that the S/N of the IFS optimized integrated spectra is
     significantly enhanced when compared with the single aperture unprocessed
     case. In some cases, the optimization based on the emission-line allows to
     characterize some of its properties better than with standard integration
     methods. It also allows to retrieve well the weak continuum features, and
     therefore to constrain better the properties of the unresolved stellar
     population. The preferred method to integrate spectra over (part of) the
     Field-of-View is ultimately dependent on the science case, and it may require
     a complex the trade off among different variables (e.g. S/N, probe area,
     spatial resolution, etc). Therefore, we provide an iterative user-friendly and
     versatile IDL algorithm that, in addition to the above mentioned method,
     allows the user to spatially integrate spectra following more standard
     procedures. This is made available to the community as part of the PINGSoft
     IFS software package.}
   {}



   \titlerunning{\shortitle}
   \authorrunning{Rosales-Ortega, Arribas \& Colina}

   \maketitle

\section{Introduction}

Large surveys of spatially integrated spectrophotometry provide a powerful means
of investigating the physical properties of galaxies at different epochs in
the history of the universe (e.g. SDSS, \citealt{York:2000p2677}; 2dFGRS,
\citealt{Colless:2001p2675}; GEMS, \citealt{Rix:2004p3993}; etc.). 
Spectral diagnostics based on integrated
optical spectroscopy can be used to constrain the star formation rate, star
formation history, stellar mass, chemical abundance, dust content, and other
main drivers of galaxy evolution \citep[e.g.][]{Tremonti:2004p1138}. 
The analysis of the integrated spectra in nearby objects allow us to compare
local and distant samples, as well as to asses the limitations in
high-redshift galaxy observations. In fact, in addition to signal-to-noise
(S/N) limitations, an incomplete spatial coverage (or aperture bias) may be
important given that many physical properties of galaxies vary depending on
the geometry and position (e.g. stellar populations, metallicity, extinction,
etc.).

The study of integrated spectral properties of galaxies poses several
observational challenges with classical spectroscopic techniques. The main
difficulty resides in obtaining an integrated spectrum of an extended object
that typically is larger than the area covered by a long slit
\citep[e.g.][]{Moustakas:2006p307}. Usually the "integrated spectrum" refers
to a single (nuclear) spectrum obtained within a fixed aperture, which
corresponds to different linear scales at different redshifts, and generally
there are limited possibilities to correct for these aperture effects.

Integral Field Spectroscopy (IFS) has become a popular observational technique
to obtain spatially resolved spectroscopy of extended objects. One by-product
of IFS data sets which was early recognised is the intrinsic capability of
integrating (by adding up some or all) the individual spectra within an
observed field or mosaic into a
single spectrum, i.e. using the Integral Field Unit (IFU) as a large-aperture
spectrograph to obtain the integrated spectra of a given Field-of-View (FoV),
increasing significantly the S/N of the final integrated spectrum. For
instance, \citet{Mediavilla:1992p3969} averaged radial profiles
from the IFS spectra to study the outwards variation of the physical
conditions and kinematics of the circumnuclear region of NGC\,4151. In a
recent example, \citet{Sanchez:2011p3844} obtained and analysed the integrated
spectrum of NGC\,628 based on the the largest IFS mosaic on a single nearby
galaxy, by adding up more than 10 thousand spectra.

Obviously, the increase of S/N in the binned or integrated spectrum
using IFS techniques has the price of a loss in angular resolution. However, a
spectrum obtained from a full 2D FoV can be better used to study the real
average spectroscopic properties of a given galaxy than those spectra taken
from different regions, or with a limited extraction aperture which recovers
only a fraction of the total optical light. This application of IFS has become
increasingly important when dealing with objects that are in a photon starving
regime, as it is usually in high-z studies, but it is also relevant when
analysing faint features in local bright objects, and in cases where the
analysis is very sensitive to S/N (e.g. stellar populations).

In principle there are several ways to combine the individual spectral of an
IFS dataset to obtain an integrated spectrum. The preferred method obviously
depends on the specific science application. For instance, if 
one wants to simulate 
a loss of angular resolution (due, for instance, to distant or
seeing effects), a simple (weighted) integration may suffice. However, in
other cases a more sophisticated methods may be preferred to obtain a trade 
between spatial information and S/N, for example the adaptive smoothing
technique \citep{Silverman:1986p4047}, which consists of correlating 2D
neighbouring information; or the Voronoi binning technique
\citep[e.g.][]{Cappellari:2003p4046,Cappellari:2009p3997} which aims to
preserve the maximum spatial resolution given a constraint on a minimum S/N
by partitioning the data using a Voronoi tessellation (see
\citealt{Diehl:2006p3996} for a comparison between Voronoi binning and adaptive
smoothing).

In this paper we describe the basis of a simple algorithm, part of the 
{\sc PINGSoft} package for IFS visualization and analysis \citep{RosalesOrtega:2011p3845},
that allows to combine IFS spectra following different prescriptions. We also
propose a method to optimize (in terms of S/N) the extraction of an integrated
spectrum from an IFS data set based on either continuum (stellar) or line
(nebular) emission features.
We show the advantages of this IFS S/N optimization method with respect to
normal fixed-aperture extractions for the integrated spectra of galaxies by
implementing the optimization to a IFS data set of nearby (Ultra)Luminous Infrared
Galaxies\footnote{LIRGs: $L_{IR} \equiv L[8-1000 \mu m]= 10^{11-12}L_{\odot}$;
  ULIRGs: $L_{IR} > 10^{12}L_{\odot}$}
[(U)LIRGs], part of the IFS low-z (U)LIRG survey \citep[][hereafter A08]{Arribas:2008p3550}.
This paper is structured as follows: in \S \ref{sec:2} we describe the S/N spectra
extraction optimization method proposed in this work,  which can be
applied to any 2D field containing spectra with continuum and/or
line emission; in \S \ref{sec:3} we apply the method to a set of IFS VLT-VIMOS
observations of (U)LIRG galaxies, describing the advantages of the optimization
by comparing the results with a fixed-aperture, single spectrum case, and by
implementing some statistical tests; the conclusions of this work are
presented in \S \ref{sec:4}. In Appen. \ref{appen:1} and \ref{appen:2} we
describe publicly available scripts in order to apply the S/N optimization
with any IFS data set.

\section{Spectra extraction optimization method}
\label{sec:2}

\subsection{S/N estimates in IFS}

Consider the case of an extended object observed within the FoV
of an IFS instrument (e.g. a galaxy or nebula), although the signal of a
particular discrete spatial element (or spaxel) may be traced on the detector in
individual exposures\footnote{In those cases where the CCD field is not
  crowded to avoid the effects of cross-talk.},
the final reduced spectrum for a particular region within the FoV of an IFS
observation is the result of a complex data processing, of either multi-exposures
and/or dithered observations, and therefore it is not straightforward to obtain
the S/N of a given spectrum in a 3D IFS datacube in a classical way, i.e. by
measuring directly the observed photon count from the CCD exposures.

Nevertheless, a {\em statistical} estimate of the S/N can be derived from the final
reduced spectrum at a given position by comparing the flux level (signal) within
a particular wavelength range with respect to the intrinsic noise of the
spectrum at the same wavelength region, i.e. the ratio of the average signal
value to the standard deviation of the signal

\begin{equation}
  \frac{S}{N} = \frac{\mu}{\sigma}.
  \label{eq:snr}
\end{equation}

This working S/N estimate will not only consider the random variations in the
detection, but all the additional sources of error introduced by the pipeline
reduction process. 
Here we are assuming that the standard deviation is dominated by noise instead
of real features, and therefore it is important in practice to select a ``clean''
spectral region for its calculation. Obviously, the prescriptions in order to
derive this estimate will vary depending on whether continuum or emission-line
features are used in order to calculate the S/N, as described below.

\subsection{Integrated spectra extraction method for optimal S/N} 

We have devised a simple spectra extraction method based on
a {\em statistical S/N} computed from emission-line features or
continuum bands. Each of the approaches would tentatively explore physically
distinct regions in the FoV of the observation, the former being consistent
with zones dominated by nebular emission (due to e.g. star formation, shocks,
LINERs, AGNs), while the latter would correspond to regions with either
dominant stellar populations or a power-law kind continuum emission.

Consider a typical 2D spectroscopy observation of an object with continuum
and/or nebular emission within the FoV of an IFS instrument. This object
--independently of the type of IFS technique-- would be sampled in discrete
spatial elements, each providing a spectrum of a particular spatial position
along a wavelength range and a given resolution depending on the instrumental
configuration during the observation. The target will be (generally)
redshifted due to the cosmic expansion; but in addition to this global shift,
the spectrum on each spaxel will show a wavelength shift due to the Doppler
effect introduced by the intrinsic velocity field of the object within the
sampled area.  The velocity field could in principle be associated to the
nebular or to the stellar component. Therefore, for the purpose of the
optimization method described below, we will assume that the observed 3D IFS
cube has been corrected from the intrinsic velocity field of the object
(although this step is optional). Note that the possibility to apply a
velocity field correction within a large FoV is one of the powers of IFS, as
this is unfeasible in long-slit spectroscopy (see Appen. \ref{appen:1} for a
description of this step).

\subsubsection {S/N estimate for the continuum}

The proposed IFS spectra selection based on the S/N is performed as follows.
Let us consider a spectrum $f=f(\lambda)$ associated with a certain spaxel $i$
of an IFS cube within an observed wavelength range $\lambda_1 < \lambda  <
\lambda_2$. 
In the case of an optimization based on the continuum, we define a continuum
band centered at $\lambda_c$ of width $w$ (in \AA) located in a relatively
{\em featureless} spectral region as the flux $f(\lambda)_c$.

We calculate a statistical-detrended S/N for this band following the definition of
\autoref{eq:snr}, i.e. the ratio of mean to standard deviation of a signal or
measurement (S/N)$_c  = \mu_c / \sigma_c$, where $\mu_c =  \bar f(\lambda)_c $
is the mean flux in the continuum band $f(\lambda)_c$ and $\sigma_c$ is the
standard deviation of the detrended flux values within that band
\footnote{The detrending consists in applying a correction for the possible
  presence of a slope within the continuum band, $\sigma_c$ is actually the 
 standard deviation of the difference between $f(\lambda)$ and a linear fit to
 $f(\lambda)$, see Appen. \ref{appen:1}.}, i.e. (S/N)$_c$ is the ratio of
the mean pixel value to the standard deviation of the pixel values over such
considered spectral range. If more than one continuum band is considered, we
can define the continuum (S/N)$_c$ of an observed spectrum as the average of
the individual S/N calculated for each of the pseudo-continuum bands. This
definition of S/N is valid when the variables are always positive (as in the
case of photon counts or flux from an emitting source) and the signal is
nearly constant \citep{Schroeder:2000p3914}, as it is the case for a detrended
featureless spectrum.

\subsubsection {S/N estimate for emission features}

In the case of the emission-line based optimization, the determination of the
{\em statistical} S/N from an observed spectrum is different from the continuum case
as the relevant $signal$ is often found in few data points with very distinct
flux values (i.e. the approximation of nearly constant signal cannot be
applied), and the $noise$ should also account for the Poisson error associated
to the flux level of the emission line. 
Therefore, we define the signal $s_{em}$ of an emission feature as the
pixel-weighted (square-root) mean of the difference between the pixel values defining the
emission line $f(\lambda)_{em}$ centered at $\lambda_{em}$ of width $w$ (in \AA),
and the mean between the two adjacent pseudo-continuum bands $f(\lambda)_{c1}$
and $f(\lambda)_{c2}$, as defined above, i.e.

\begin{equation}
  s_{em} = \frac{1}{\sqrt N}  \sum\limits_\lambda  {\left[ {f(\lambda)_{em}  - \left\langle
        {f(\lambda)_{c1} , f(\lambda)_{c2} } \right\rangle } \right]},
\end{equation}

\noindent 
where $N$ is the number of pixels within $w$.
The reason of the wavelength index running in both the emission line feature
and the adjacent pseudo-continuum bands is to account somewhat for any
possible tilt on the continuum shape at the relevant spectral region, as the
detrending applied in the continuum case.

We consider two sources of noise $n_{em}$ in the case of an emission-line feature,
the first one related to the intrinsic noise of continuum region where the
emission line is found, calculated via the mean between the standard deviation
of the flux values within the two adjacent pseudo-continuum bands
$\sigma(\lambda)_{c1}$ and $\sigma(\lambda)_{c2}$ (in a similar manner as in
the S/N continuum case), while the second one is related to the
Poisson noise associated to the flux level of the emission line. The latter
is derived by generating a $N$-element array of random numbers drawn from a
normalized Poisson distribution
\footnote{Using the formula of section 7.3 in {\em Numerical Recipes in C: The
  Art of Scientific Computing}, \citep{Press:1992p3916}.} 
multiplied by the square root of the observed flux $f(\lambda)_{em}$, i.e.

\begin{equation}
  n_{P} = P(k) \times \sqrt {f(\lambda)_{em} }.
\end{equation}

\noindent
The standard deviation of the resulting array, $\sigma_P$, is added in quadrature in order to
obtain the total noise related to the emission-line feature $n_{em}$, i.e.

\begin{equation}
  n_{em} = \sqrt { \left\langle {\sigma_{c1} ,\sigma_{c2} } \right\rangle ^2 + \sigma_{P} ^2 },
\end{equation}

\noindent
the S/N of the emission-line feature is therefore given by

\begin{equation}
S/N_{em} = \frac{s_{em}}{n_{em}}.
\end{equation}

\noindent
As a sanity check and for comparison purposes, the {\em statistical} S/N
defined above was calculated for long-slit spectra of nearby extragalactic
regions with S/N derived from the CCD photon count ranging between $\sim
10-50$ in the continuum and $\sim 10-200$ in the \ha\ emission line
$\lambda_{em} = 6563$ \AA, obtaining a good agreement between the $real$ and
the {\em statistical} S/N (the differences being of the order of $10-20
\%$). Note that an absolute agreement is not compulsory, since the (S/N)$_c$
and (S/N)$_{em}$ are used in the optimization method as a guideline in order
to discriminate between high/low S/N spectra.

\subsection {Optimization procedure} 

\begin{figure}
  \centering
  \includegraphics[width=\hsize]{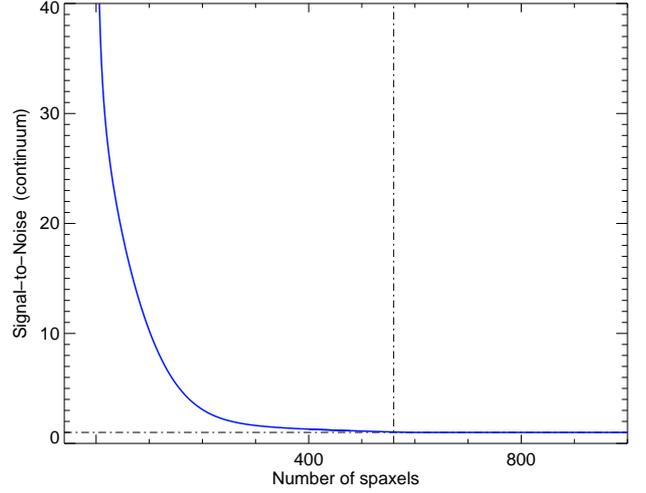}
  \caption{
    Common behaviour of an IFS spectral cube sorted in decreasing order of the
    S/N. 
    The figure corresponds to a VIMOS-IFU cube section containing 1000 spaxels of an IFS
    observation of the nearby LIRG IRAS F06295-1735 with continuum emission.
    The vertical axis shows the S/N calculated on a relatively {\em featureless}
    narrow continuum band at $\lambda_c = 6200$ \AA\ for each spaxel of the 3D
    cube, while the horizontal axis corresponds to the spaxels ordered by
    decreasing (S/N)$_c$.
    Few spaxels show high S/N (i.e. $> 10$), decreasing rapidly to
    values in which the continuum is dominated by noise ($S/N_c \sim 1$).
    \label{fig:snr}
  }
\end{figure}

The actual optimization method is performed by first calculating and recording the
(S/N)$_c$ and (S/N)$_{em}$ for each of the spectra in a velocity-corrected 3D cube
following the above prescriptions. Then the spaxels are sorted in two groups
(continuum and emission-line),
one in decreasing order of the (S/N)$_c$ value and the other by decreasing order
of the (S/N)$_{em}$. \autoref{fig:snr} shows the typical behaviour of an IFS
spectral cube sorted in this manner, the vertical axis shows the (S/N)$_c$ and
the horizontal axis corresponds to the spaxels ordered by decreasing (S/N)$_c$.
In this particular case, the figure corresponds to a cube section of an IFS
observation in the optical ($\sim5200-7000$ \AA) of the nearby luminous
infrared galaxy IRAS F06295-1735 (ESO 557-G002) containing 1000 spaxels with
continuum emission \citepalias[][see \autoref{sec:3}]{Arribas:2008p3550}.
Few spaxels show a relatively high S/N, decreasing rapidly to values in which
the continuum is dominated by noise. In the next step, the optimization method
neglects all those spaxels with S/N $\approx 1$ in both groups, creating two
different spectral samples. 
In \autoref{fig:snr}, this threshold is found around the sorted
spaxel number 550 in the case of the continuum-based S/N optimization,
i.e. $\sim$ 900 spaxels are dismissed from the original $40 \times 40$ spaxels
datacube\footnote{Note
  that this ordering has nothing to do with the spatial position or
  the data format of the IFS cube}. Visual inspection of the rejected spaxels
shows that these are consistent with spaxels sampling the sky (null continuum)
or regions with very low signal in the continuum and/or strong background
noise along the entire spectral range.

At this point, two groups of spectra with S/N $> 1$ have been selected. The
process can be continued by integrating:

\begin{enumerate}
\item[i)] All the spectra over a specific geometrical region of the galaxy/FoV
\item[ii)] All the spectra above a given S/N threshold
\item[iii)] A combination of i) and ii), or
\item[iv)] All the spectra that optimize the S/N over a given region
\end{enumerate}

\noindent
Methods i), ii) and iii) are conceptually trivial, and therefore we focus now on in
this section on the proposed method for iv), which as i)-iii) can be applied
for the (S/N)$_c$ or (S/N)$_{em}$, with or without the velocity field
correction.
The actual implementation is the following, the method creates an integrated
spectrum for each of the spaxels subsample by adding the spectra in the sorted
order as described before, in each step the procedure calculates the S/N in the
continuum and/or emission line feature in the new generated spectrum, e.g. in
the first step, an integrated spectrum is created by adding the spectrum with
the highest S/N to the 2nd highest S/N spectrum on each group, a S/N is
calculated and recorded for this integrated spectrum; in the next step, the
3rd S/N-ranked spectrum is added to the previous one, creating a new
integrated spectrum on which a new set of S/N values are calculated, and so on.
The rationale of this criterion is the following: spectra of good quality (high
S/N) should contribute to enhance the S/N of the integrated spectrum on each of
the two groups, if the inclusion of a new spectrum (spectra) decreases the
S/N of the integrated spectrum with respect to the values found in previous
steps, then the individual S/N of the spectrum (spectra) where this 
{\em turn-off} is found marks a tentative threshold value (or range) sought
for high quality spectra within an IFS cube.

\begin{figure}
  \centering
  \includegraphics[width=\hsize]{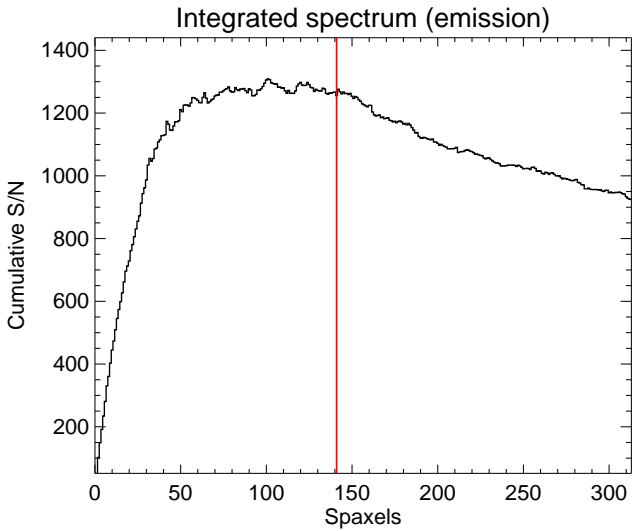}
  \caption{
    Cumulative S/N calculated on an emission-line feature (H$\alpha$) from the integrated
    flux obtained by adding the spectra in the sorted order as shown in
    \autoref{fig:snr} for the IFS observation of IRAS F06295-1735.
    High S/N spectra contribute to enhance the S/N of the
    integrated spectrum until a point in which the inclusion of additional
    (low-S/N) spectra does not contribute to increase the overall S/N of the
    integrated spectrum, creating either a {\em plateau} or a {\em turn-off},
    marking the threshold value (range) sought for high quality spectra within
    an IFS cube (e.g. red vertical line).
    \label{fig:integ}
  }
\end{figure}

\autoref{fig:integ} shows the {\em cumulative} S/N calculated on the \ha\
emission-line of the integrated spectrum obtained by adding
subsequently the S/N-sorted spectra of the IFS observation of IRAS
F06295-1735 as shown in \autoref{fig:snr}. For the first 50 spaxels or so, the
S/N raises substantially as consecutive spectra are added to the integrated
spectrum, until a point in which the inclusion of additional spectra 
creates a {\em S/N plateau} around $S/N_{em} \sim 1300)$. After $\sim 150$
spaxels, the S/N of the integrated spectrum shows a {\em turn-off} in which
the values start to decline as we add more spectra, marking the tentative
threshold value (range) sought for high quality spectra within the
IFS cube. In an (tentative) automated procedure, this threshold might be simply taken as
the S/N of the spaxel in which we find the maximum of the curve shown in 
\autoref{fig:integ}. 
The shape of the this curve obviously depends on both the target intrinsic
characteristics and the observations conditions. The presence of a {\em
  plateau} in the case shown in \autoref{fig:integ} illustrates that in
practice the optimization may be more complex task, and require some kind of
interactive process. Even when optimal S/N is the main criterion for
obtaining an integrated spectrum, in practice there may be a range of
alternative options. 
As one of the advantages of IFS is to have 2D information, the
final selection of a sample of spaxels has to be a trade-off between regions
with S/N relevant for the intended scientific analysis and a sufficient
number or spaxels in order to recover as many possible regions of the target
within the FoV of the instrument. The inspection of the trends as
shown in \autoref{fig:integ} can help in order to select either qualitatively
(by the trend of the curve vs. spatial coverage) or quantitatively (by
the point of maximum S/N or desired range if signal) a subsample of spaxels.
For instance, on the light of \autoref{fig:integ}, it may be scientifically
preferable the selection of only the 75 spaxels with highest S/N or,
alternatively, extend this selection to 150 spaxels which will probe a larger
area but still maintaining the global S/N. This example illustrates that the
actual process for optimizing an integrated spectra may be complex, and
requires interactive user-friendly versatile tools to carry it out in an
efficient way.

\begin{figure*}
\sidecaption
\includegraphics[width=12cm]{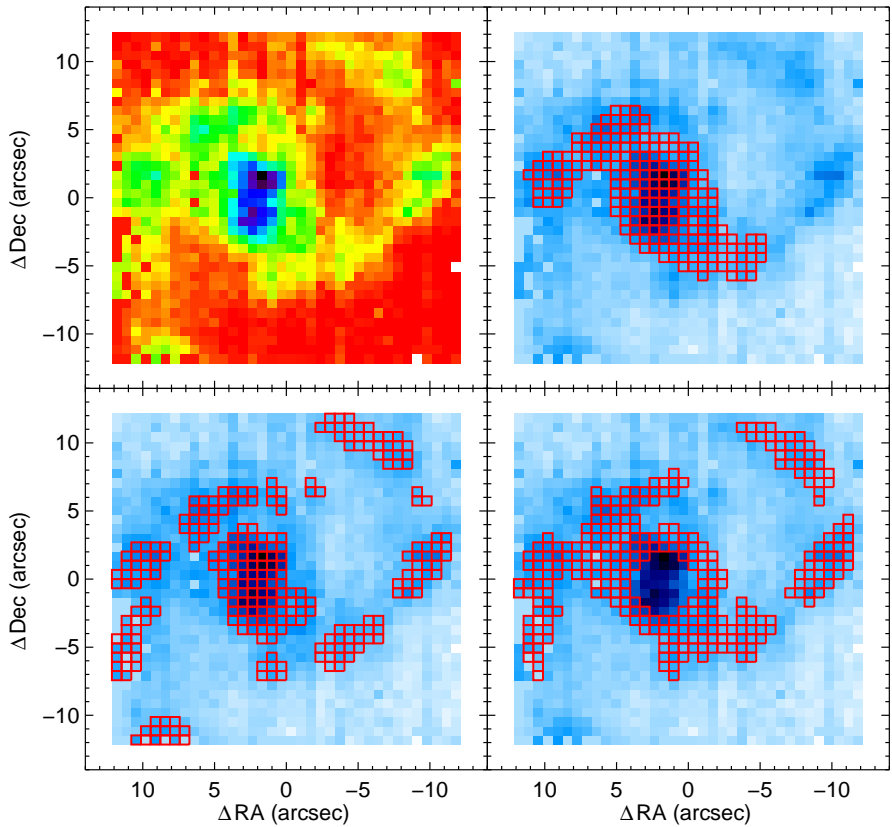}
 \caption{
    Visual example of the S/N optimization method applied to the VLT-VIMOS
    observation of the LIRG IRAS F06295--1735. 
    Top-left panel: $H\alpha$ ``narrow-band'' image showing the
    morphology of the barred spiral galaxy. The colour intensity of each
    spaxel is scaled to the total flux intensity for that particular region
    which includes H$\alpha$ + [NII] line-emission and continuum.
    Top-right: selected spaxels using the S/N optimization continuum
    criterion. Bottom-left: selected spaxels after
    applying a S/N optimization on the H$\alpha$ emission-line. Bottom-right:
    continuum S/N selection after discarding the nuclear spaxels in a
    circular aperture of 2~kpc.
    \label{fig:vimos}
  }
\end{figure*}

\autoref{fig:vimos} shows a visual example of the implementation of the S/N
optimization method described above (\autoref{fig:snr} \& \autoref{fig:integ}) 
to a VLT-VIMOS observation of the nearby LIRG IRAS F06295-1735
\citepalias{Arribas:2008p3550}. The top-left panel shows a
``narrow-band'' image extracted from the IFS datacube centered at \ha\ with a
width of 100 \AA. The panel shows that the morphology of the galaxy corresponds to a
barred spiral with strong emission at the nucleus. Previous studies have shown
that the line ratios at all locations in the galaxy are consistent with
photoionization by stars \citep{RodriguezZaurin:2011p3970}. This target
represents a good example of a IFS observation where we can expect spatially
separated regions with different S/N in the continuum and line emission.
For each spaxel, the S/N in the continuum was calculated at a relatively 
{\em featureless} spectral narrow-band of 50 \AA\ width at $\lambda_c =
6200$ \AA, while the S/N of the emission-line
was derived using the \ha\ line at $\lambda_{em} = 6563$ \AA. The spaxels for
each subsample were sorted (e.g. \autoref{fig:snr}) and a cumulative S/N was
derived for the integrated spectrum by adding consecutive S/N-ranked spaxels
(\autoref{fig:integ}). The S/N threshold value for each subsample was chosen
in this exercise as the one in which the S/N of the integrated spectrum
has a {\em turn-off} after the {\em plateau} of high S/N values (e.g. the
red vertical line in \autoref{fig:integ} in the case of the emission-line
subsample).

In the top-right panel of \autoref{fig:vimos}, the selected
spaxels after applying the S/N optimization are drawn with red squares, and
they sample spatially the nucleus and most of the bar, but rejects as low
S/N regions those spaxels corresponding to the spiral arms. 
The bottom-left panel corresponds to the spaxel selection using the S/N
criterion on the \ha\ emission-line, in this case the selected spaxels clearly
sample the nucleus of the galaxy and some knots and individual regions along
both the spiral arms, as it could be expected. 
Interestingly, the selected spaxels via the \ha\ emission-line do not coincide
with those in the continuum case. Note that some spaxels are
selected in both methods (e.g. at the nucleus) but that in general they sample
clearly distinct regions of the galaxy. Depending on the scientific case, one
can be interested in extracting one or another subsample, a combination of
them, or a trade S/N vs. geometrical region probed. For instance, in the
bottom-right panel of \autoref{fig:vimos}, the
nuclear region of the galaxy was discarded in the S/N optimization by applying
a circular aperture of 2~kpc in physical scale; the rest of spaxels subject to
a S/N optimization on the continuum as before, a new S/N growing-curve of
the extra-nuclear regions was derived and a threshold value was chosen using
the same criteria as before. In this case, the selected spaxels trace the bar,
most of the north spiral arm and some regions of the south arm, following the
morphology shown in the top-left panel of the same figure. This example shows
that the proposed S/N optimization can be very advantageous on studies
of the global properties of galaxies (or other targets) by making use of 
spatially resolved IFS in order to obtain distinctive integrated spectra 
from regions dominated by either continuum and/or line emission, depending on
the posterior intended analysis.

\begin{figure*}
  \centering
  \includegraphics[width=0.47\hsize]{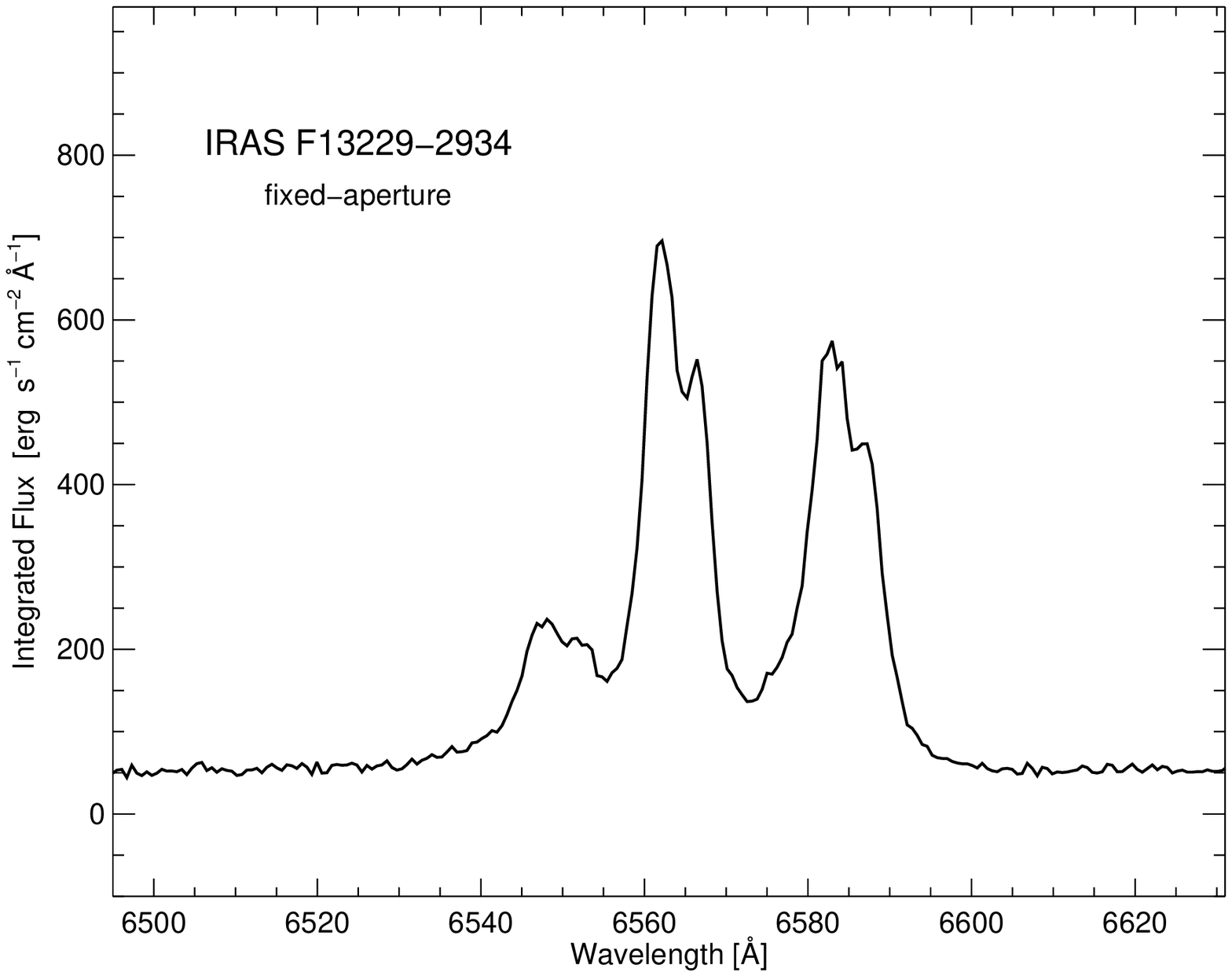}
  \includegraphics[width=0.47\hsize]{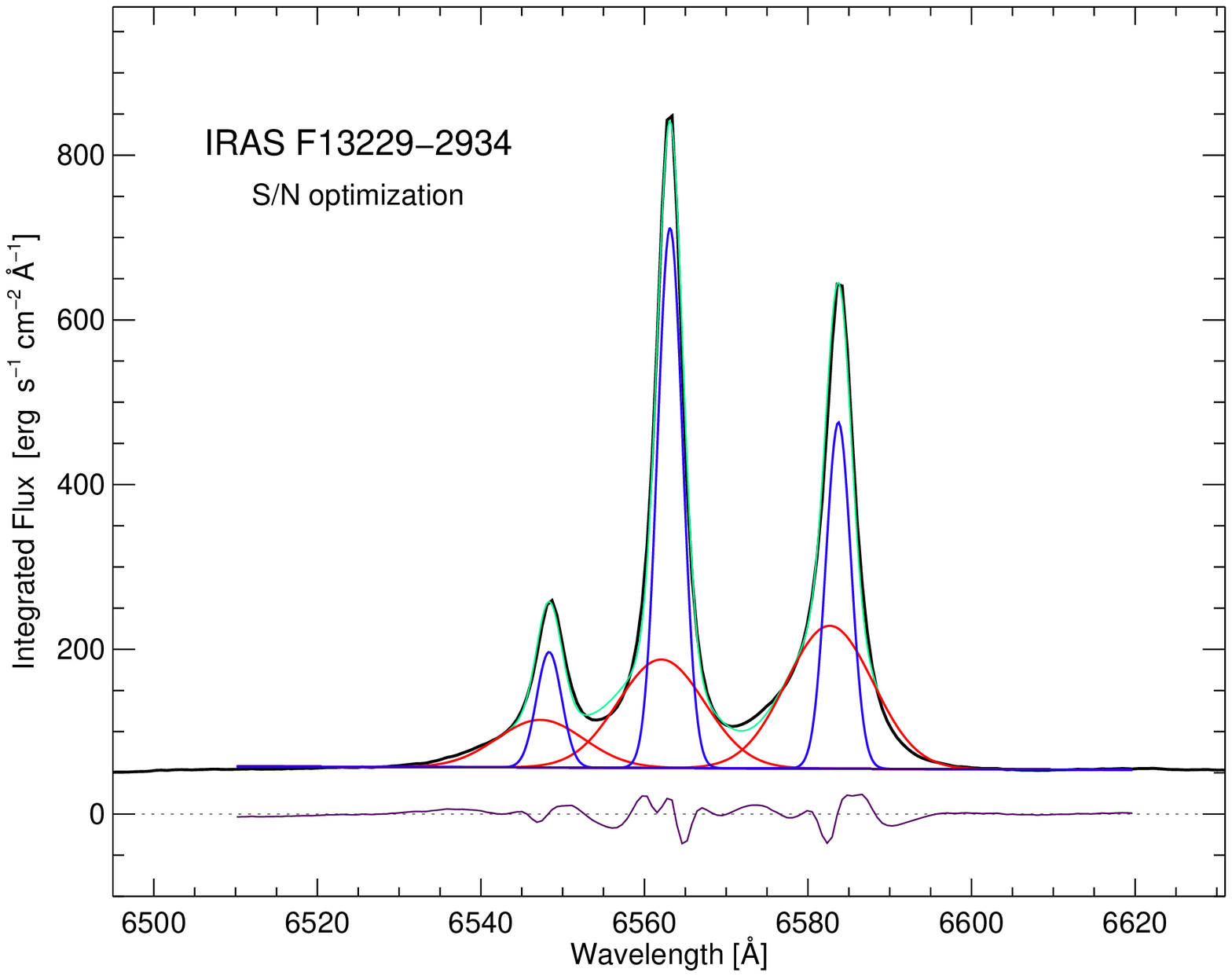}
  \caption{
    {\em Left}: Integrated spectrum of the LIRG IRAS F13229-2934 obtained by
    adding the IFS spaxels within a simulated 10 arcsec circular aperture
    without any velocity-field correction. {\em Right}: The black-line shows
    in black the integrated spectrum of the same galaxy obtained by applying
    the S/N optimization method using the emission line criterion on the
    H$\alpha$ line. The blue and red lines correspond to the fit of a narrow and
    broad components respectively, while the the green line stands for the total
    fit. The residual between the observed spectrum and fit models is shown
    at the bottom of the panel.
    \label{fig:fits}
  }
\end{figure*}

Experience with the optimization technique shows that the the final
extraction depends little on the configuration parameters, as long as they are
actually used as a proxy of "featureless continuum" and that the pixels
defining the emission feature contains exclusively the emission line proxy
(i.e. avoiding "blending wings" like in the \ha\ $ + $ \nii\ system), and
that the pseudo-continuum bands sample real regions without strong absorption or
emission features. Moreover, the results are stable if the parameters are
slightly modified (e.g. changing the central wavelength or width by 5-10 \AA),
but the user might expect variations on the results if these parameters change
significantly, and of course depending on the nature of the data.

\subsection {Integration options and routines description}

Although the optimization method here proposed is conceptually simple, its
actual implementation in real IFS data might not be straightforward. 
Therefore, we have produced a series of IDL scripts that can be applied to
basically any IFS datacube in order to implement the S/N optimization here
described. They form part of the {\sc PINGSoft} package for IFS visualization,
manipulation and analysis \citep{RosalesOrtega:2011p3845}, a software
optimized for large databases and a fast visualisation rendering. 
The routines can perform a S/N optimization on the basis of a (S/N)$_c$ or a
(S/N)$_{em}$, and with and without the velocity field correction.
This allows a high degree of versatility to obtain an integrated spectrum for
an optimal scientific use. The interactive user-friendly and versatile nature
of the tool allows one to carry out this process in an efficient way.

The relevant routines are named {\tt vfield\_3D.pro}, {\tt s2n\_ratio\_3D.pro},
and {\tt s2n\_optimize.pro}. Note that they are not standalone codes, the whole 
{\sc PINGSoft} library should be present for a proper execution.
The {\tt vfield\_3D.pro} performs a wavelength shift by applying a
cross-correlation with respect to a reference spectrum on those targets with
intrinsic velocity fields likely to be large (e.g. U-LIRGs).
The {\tt s2n\_optimize.pro} routine performs the S/N optimization described in
this paper, while the {\tt s2n\_ratio\_3D.pro} code extracts interactively
those spaxels above a given S/N threshold.
Note that the routines are independent, and the application of the 
{\tt vfield\_3D.pro} code is optional.

Following the philosophy of {\sc PINGSoft}, these codes are user-friendly,
robust and easy to implement. A proper combination of the available
parameters (see the Appendices) allows to extract those regions above a
certain S/N in the continuum and/or emission (on either corrected or not
corrected velocity-field spectra), and to optimize the S/N on the extracted
spectra on both the emission and continuum, obtaining as by-products the
statistics and distribution of the S/N for all the spaxels, 2D S/N maps for
the continuum and emission, graphical outputs, and the extracted spectra for
both samples in FITS format ready to be easily visualised and analysed.
A more detailed description of the scripts and a working example on their
implementation can be found in Appen. \ref{appen:1} and \ref{appen:2}. The
{\sc PINGSoft} package is freely available at project
webpage\footnote{\url{http://www.ast.cam.ac.uk/ioa/research/pings}}, under the
Software section.

\section{Potential of the optimization method}
\label{sec:3}

\subsection {Test Sample}

In this section we show some examples of how the implementation of the S/N
optimization can result very advantageous in order to improve substantially
the quality of potentially analyzable data. For that purpose, we applied the S/N
optimization to an optical ($5250-7450$ \AA) IFS data set of nearby (U)LIRGs
using the VLT-VIMOS instrument \citep{LeFevre:1998p3053}. 
A detailed description of the observations and data reduction can be found in
\citetalias{Arribas:2008p3550}.

Once the data are reduced, the \ha\ line was used as a
reference spectrum in order to perform an intrinsic velocity field correction
by applying a wavelength cross-correlation to all the spaxels in the
individual cubes.
S/N-optimized integrated spectra was extracted for all the targets from the
velocity-corrected cube, using a continuum narrow-band centered at 
$\lambda_c = 6200$ \AA\ (rest-frame) with a 50 \AA\ width in the case of the
S/N-continuum optimized sample and $\lambda_{em}$ = \ha\ $=6563$ \AA\ in the
case of the S/N line-emission optimized spectra.
The spectral region in the continuum was selected since at those wavelengths
the spectra does not show strong absorption/emission features, so that the
condition of nearly-constant signal is fulfilled. On the other hand, the \ha\
line emission is the most prominent emission feature within the spectral range
of the data.
A detailed analysis and/or comparisons between the physical properties derived
from each set of spectra are beyond the scope of this paper, here we just
focus on the advantages and applications that high-S/N optimized integrated
spectra obtained from a 2D IFS cube can provide when applying common and
specific analysis techniques on both the stellar and nebular components of the
integrated spectrum of a galaxy.

\subsection {Emission S/N optimized spectra test} 

The left-panel of \autoref{fig:fits} shows the rest-frame \ha\ +
\nii\ spectral region of the integrated spectrum of the LIRG IRAS F13229-2934
obtained by adding the IFS spaxels within a simulated 10 arcsec circular
aperture (73 spaxels), no velocity correction was applied. The complex
emission line profiles are clearly the result of several kinematical
components mixed in the integrated spectrum, which presents some level of
noise in the continuum as the consequence of the mixing of spectral features
due to the velocity field of the object and the inclusion of regions with
significant noise within the simulated fixed-aperture. On the other hand, the
right-panel of \autoref{fig:fits} shows in black the integrated spectrum of IRAS
F13229-2934 obtained by applying the S/N optimization method using the
line-emission line criterion on the \ha\ line (64 spaxels). In this case, as
the selected spaxels are corrected individually by the intrinsic velocity
field of the object, the emission lines are consistent with a single narrow
Gaussian profile plus a well identified blue-shifted broad component.

Given the high S/N of the optimized spectrum and cleaner shape of the
emission-line profiles, we are in the position to 
study the average local kinematic properties associated to these two
components (i.e. narrow and broad). The blue and red lines drawn on the
right-panel of \autoref{fig:fits} correspond to the fit of a narrow and broad
components respectively, while the the green line corresponds to the total
fit, i.e. the sum of the components.

Note that the implementation of the S/N optimization has allowed to better
decouple the overall kinematic components of the integrated spectrum of the
galaxy, and to recover more line-emission within the FoV of the IFS
observation than the single, fixed-aperture case.
A by-product of the determination of the different kinematic
components is the correct determination of the total line-emission in a given
object. In general, a single-line Gaussian fit to the integrated spectrum of 
an object with an important broad component would underestimate the total flux
of the emission lines due to the presence of line-wings.
Take as an example the \nii/\ha\ line ratio commonly employed when deriving
ionization conditions \citep[e.g.][]{MonrealIbero:2010p3872} or nebular abundances
\citep[e.g.][]{Denicolo:2002p361},
if the flux for each line species is the sum of a narrow plus a broad
component we find important deviations from the single-line case.
This fact can be specially important when dealing with regions of high
excitation, turbulence, AGNs or high-z objects.

\subsection {Continuum S/N optimized test}

\begin{figure}
  \centering
  \includegraphics[width=\hsize]{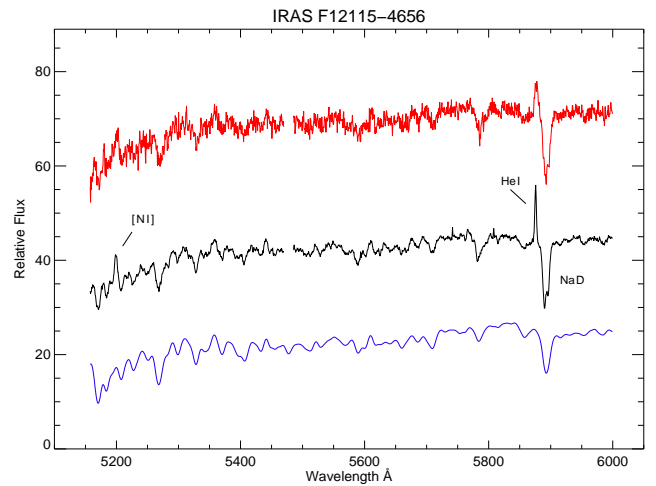}
  \caption{
    Close-up of the integrated spectrum of the LIRG IRAS F12115-4656. The
    top-red line corresponds to the spectrum obtained by integrating the VIMOS
    spaxels within a simulated circular fixed-aperture of 10 arcsec centered
    at the nucleus of the galaxy in the IFS data cube without any velocity-field
    correction. The central spectrum in black corresponds to the S/N {\em
      continuum}-optimized integrated spectrum of the same galaxy, some
    spectral features are labelled for an easy identification.
    The blue spectrum at the bottom corresponds to a simple SSP synthetic model
    fitted to the S/N-optimized integrated spectrum, as described in the text.
    All spectra are displayed  in a relative flux scale.
    \label{fig:spec}
  }
\end{figure}

Another advantage of the S/N optimization method resides in the retrieval of
weak continuum features in the high-S/N, velocity-field corrected integrated
spectra. \autoref{fig:spec} shows a close-up on the spectral region 
$\sim 5200-6000$ \AA\ of the integrated spectrum of the
LIRG IRAS F12115-4656. The top-red line corresponds to the integrated spectrum
obtained within a simulated circular fixed-aperture of 10 arcsec centered at
the nucleus of the galaxy in the IFS data cube without any velocity-field
correction, while the central spectrum in black corresponds to the 
{\em continuum}-optimized integrated spectrum of the same galaxy, in a
relative flux scale, some spectral features are labelled for an easy
identification. Visual comparison between both spectra shows clearly the
differences between the fixed-aperture and the S/N-optimized integrated
spectra, the former showing a much higher level of noise, while on the latter
the details of the stellar absorption features on the continuum can be clearly
seen.

Currently, one of the state-of-the art techniques to analyse the unresolved
stellar populations of galaxies is the {\em spectral synthesis} of an
integrated spectrum, which consist in the decomposition of an observed
spectrum in terms of a combination of single stellar populations (SSP) of various
ages and metallicities, producing as output the star formation and chemical
histories of a galaxy, together with its extinction an velocity
dispersion \citep[e.g.][]{CidFernandes:2005p357,CidFernandes:2010p3859,Koleva:2009p3414,MacArthur:2009p3412}.
This is achieved by a full spectrum fitting including the continuum shape and
absorption features. Several authors have developed different fitting codes,
either based on spectral indices or full spectrum modelling, and there is a
plethora of literature which discuss the pros and cons of the different techniques
\citep[e.g. see][and references therein]{MacArthur:2009p3412,Walcher:2011p3994}.
Independently of the fitting technique, among the requirements for the
effective derivation of the physical properties of galaxies using these
methods we require a good spectral resolution, a wide spectral range
{\em and} a sufficient S/N of the observed spectrum. 
The high S/N and the detailed structure of stellar features found in the optimized
integrated spectra of galaxies (as shown in the middle spectrum of
\autoref{fig:spec}), can contribute significantly to recover the physical
properties of the unresolved stellar populations in galaxies in a more
efficient way.

To illustrate this point we performed the following exercises.
The blue spectrum at the bottom of \autoref{fig:spec} corresponds to a very
simple SSP synthetic model fitted to the S/N-optimized integrated spectrum of
IRAS F12115-4656 using the {\sc FIT3D} spectral fitting procedure after
\citet{Sanchez:2007p1696}.
The fitting was performed using a linear combination of a grid of six SSP
templates from the {\sc Miles} spectral library after \citet{SanchezBlazquez:2006p3860}
with a very young, intermediate and old population (0.1, 1 and 17 Gyr) and two
extreme metallicities ($Z=0.004$ and 0.02).
The templates were first corrected for the appropriate systemic velocity and
velocity dispersion (including instrumental dispersion).
A spectral region of $20-30$ \AA\ width around each detected emission line was
masked prior to the fitting, including also the regions around the sky-lines.
The purpose of this simple fitting was merely to compare qualitatively the
shape and structure of the continuum and stellar features between the
synthetic model and the optimized integrated spectrum, we do not intend to
derive the luminosity or mass-weighted age, metallicity
or dust content of the galaxy, as we are aware that the spectral range of the
VIMOS (U)LIRGs data is not long enough to break well-known degeneracies
\citep{MacArthur:2009p3412,Sanchez:2011p3844}.
Note that, even with this very simple fitting scheme employing only a few SSP
templates, the agreement between the structure of the stellar features in the
optimized integrated spectrum and the synthetic model are very good
(excluding the emission-line features on the observed spectrum).

\begin{figure}
  \centering
  \includegraphics[width=\hsize]{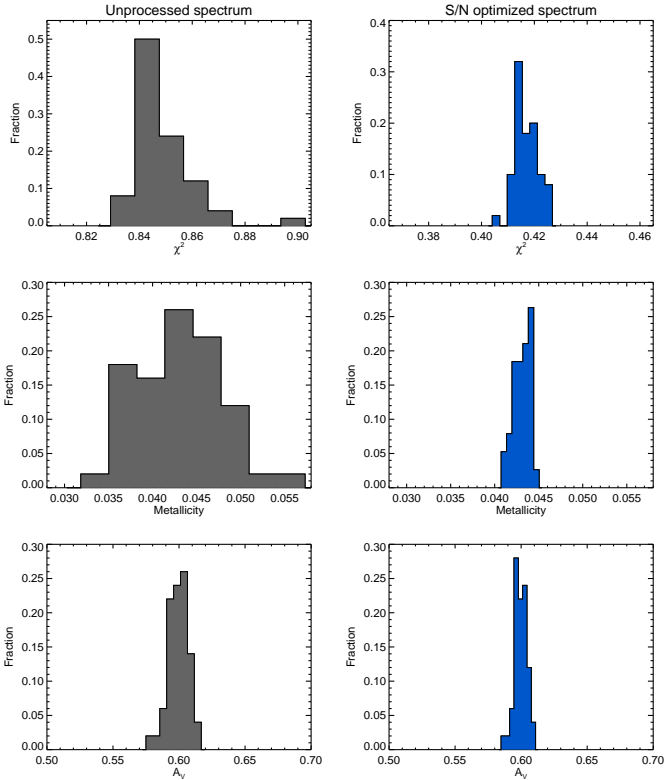}
  \caption{
    Histograms of the $\chi^2$-fitting, metallicity $Z$ and dust
    attenuation $A_V$ distributions obtained from the SSP fitting to 100
    simulated spectra based on the {\em noise-residual}, as explained in the
    text. The left-column panels correspond to the fixed-aperture, unprocessed
    spectrum and the right-column panels to the S/N optimized spectrum of IRAS
    F12115-4656, as shown in \autoref{fig:spec}.
    \label{fig:mc}
  }
\end{figure}

A more quantitative way of exploring the benefits of having a S/N optimized
spectrum for a spectral synthesis is to analyse, in statistical terms, how the
fitting technique performs with an unprocessed spectrum compared to an
optimized one. For that purpose we considered the
unprocessed, fix-aperture integrated spectrum (240 spaxels) and the S/N-optimized integrated
spectrum (206 spaxels) of IRAS F12115-4656 shown in \autoref{fig:spec}. Each spectrum was
fitted using {\sc FIT3D} as described above, but with a different grid of six
SSP templates with a fixed age (2.5 Gyr) and a full
range of metallicities ($Z=0.0001, 0.0004, 0.004, 0.008, 0.02, 0.05$).
We chose to fix one of the fitting parameters as we are only interested in
exploring statistically how the S/N-optimized spectrum can contribute to
improve the fitting technique (by comparing the behaviour of the other two
free parameters).
Once we derived a first synthetic approximation of the underlying
stellar population for each spectrum (unprocessed and S/N-optimized), this was
subtracted from the original spectrum to obtain a {\em residual} pure
emission-line spectrum. The emission lines were fitted to a single Gaussian
function per emission line plus a low order polynomial function for the local
continuum and a pure gas-emission spectrum model was created, based on the
results of the last fitting procedure, using only the combination of Gaussian
functions. The pure gas-emission model was then subtracted from the original
observed spectrum to produce a gas-free spectrum. This spectrum was fitted
again using the same combination of SSPs, as described before (but without
masking the spectral range around the emission lines), deriving a new,
synthetic SSP model of the composite stellar population
\citep[see][for a detailed explanation of this fitting technique]{Sanchez:2011p3844}.

The iterated synthetic model was then subtracted again to the corresponding
observed spectrum (fix-aperture and optimized spectrum), creating a 
{\em noise-residual} on each case. Given the difference in the intrinsic
noise level and structure of the stellar features between the optimized and
unprocessed spectra (\autoref{fig:spec}, the {\em noise-residual} of the
unprocessed spectrum presents a higher RMS (7.28) value with respect to the residual
of the optimized spectrum (2.98), as expected. This is in principle one way of
measuring quantitatively the difference on the fitting between the two spectra.
However, as we want to explore statistically the impact of one spectrum with
respect to the other in the fitting technique, we performed a Monte Carlo
simulation using as an input the observed spectrum and adding a simulated 
{\em noise-spectrum} generated from the {\em noise-residual} described above.
For doing so, we assumed that the {\em noise-residual} is a good measure of
the intrinsic noise of the observed spectrum, and we applied a median filter
along the {\em noise-residual} with a width equal to the FWHM of the
instrumental resolution (in order to account for the intrinsic correlation of
the noise). Then, each pixel of the {\em noise-spectrum} was
randomly drawn from a Gaussian distribution with a mean equal to the
corresponding pixel value of the filtered {\em noise-residual}. We simulated
100 spectra for each input spectrum (unprocessed/optimized + {\em noise-spectrum}),
and for each spectrum of the simulated data we applied the iterative SSP
fitting technique described above, with the same grid of SSP templates as
before, deriving and recording for each spectrum the $\chi^2$ of the fitting
and the values of the luminosity-weighted metallicity and dust content of the
composite stellar population.

\autoref{fig:mc} shows the histograms of distributions for the set of values
derived for each of the two observed spectra, the left-column corresponds to
the fixed-aperture, unprocessed spectrum and the right-column to the S/N
optimized spectrum of IRAS F12115-4656. The top-panels correspond to the 
$\chi^2$ value of the fitting, the middle-panels to the metallicity $Z$ and the
bottom-panels to the dust content in terms of the visual extinction $A_V$.
As mentioned before, we are not interested in the absolute (physical) values
of the free parameters, but on their relative differences as measured by the
same fitting technique and templates grid for the distinct set of fitted
spectra. In the case of the unprocessed spectrum, the distribution of the
$\chi^2$ has a mean of 0.846 and a standard deviation 0.012, while in the case
of the optimized spectrum the mean is found at 0.417 with a standard deviation
0.004, i.e. the mean $\chi^2$ value of the fitting is more than a factor of
two higher for the unprocessed spectrum compared to the optimized one (with
three times higher dispersion). In the case of the metallicity, the mean $Z$
in both distributions is very close ($Z \approx 0.0430)$, as it can be
expected given the sort of simulation performed, but the dispersion of both
distributions is complete different, in the case of the optimized spectrum the
standard deviation in $Z$ is 0.0009, while for the unprocessed spectrum this
is 0.0048, i.e. a factor of 5 higher. The same effect occurs for the total dust,
both distributions have a mean close to $A_V \approx 0.599$, the main
difference being on the their standard deviation, that in the case of the
optimized spectrum is 0.0046, while for the unprocessed one is 0.0074,
i.e. nearly a factor of two higher. The lower dispersion in the dust
attenuation can be explained by the fact that the spectral range of the
VIMOS (U)LIRGs observations is not very sensitive to changes in the
extinction, and that the main difference is found on the metallicity precisely
because the overall stellar features in the optimized spectrum have a higher
level of detail and less intrinsic noise than the unprocessed spectrum.
Although the results of this simple simulation should be taken with caution
(as the derived values depend strongly on the chosen free parameters, the
fitted spectral region, input spectrum and the grid of SSP templates), this
exercise shows how for the S/N-optimized spectrum the SSP fitting technique
reduces largely the uncertainties in the derived physical parameters of the
stellar populations.

\subsection{Comparison with other techniques}
\label{sec:vor}

\begin{figure*}
\sidecaption
\includegraphics[width=12cm]{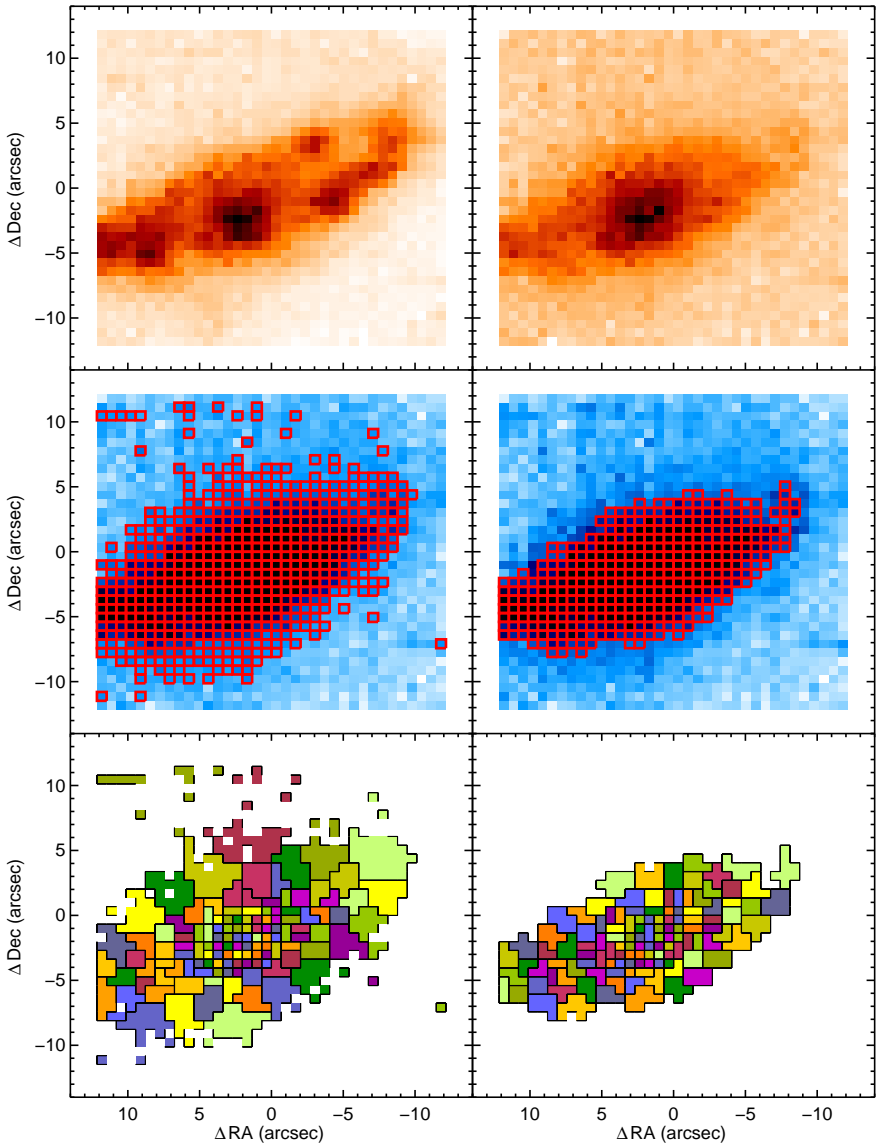}
 \caption{
   Comparison of the S/N optimization method with other techniques using the
   VLT-VIMOS observation of IRAS F10409-4556.  
   Top-panels: $H\alpha$ narrow-band (left) and V-band continuum (right)
   images showing the morphology of the galaxy. Middle-left panel: selected
   spaxels using a simple S/N threshold cut ($> 2$) on the continuum at 5700
   \AA\ (100 \AA\ width). Middle-right panel: selected spaxels applying the
   S/N optimization method in the continuum using the same parameters as in
   the S/N threshold case. Bottom-panels: Voronoi binning of the selected
   spaxels using the S/N threshold technique (left) and the S/N optimization
   (right), with a target minimum S/N within each bin of 5.
    \label{fig:vor}
  }
\end{figure*}

In this section we compare the S/N optimization with other techniques in order
to better quantify the actual gain of the presented method. \autoref{fig:vor}
shows the VLT-VIMOS observation of the spiral LIRG IRAS F10409-4556
(\citepalias{Arribas:2008p3550}), the upper panels show the galaxy as seen in
\ha\ emission (left) and in the V-band continuum (right). In the case of the
continuum image we note that the signal drops smoothly from the nucleus of
the galaxy to the outer parts where the signal level is comparable to the
background, making difficult to discriminate the boundaries of the object. A
typical way of selecting those spaxels to which perform a subsequent analysis
is to apply a S/N cut above a given threshold. This is shown in the
middle-left panel of \autoref{fig:vor}, the spaxels drawn in red correspond to
those regions above a S/N threshold of 2 as calculated in the continuum at
5700 \AA\ (100 \AA\ width). The selected spaxels cover the morphology of the
galaxy as seen in the upper panels, but the boundaries are not well defined
and most importantly, the selected regions include spaxels not related to the
object that are spread around the FoV of the IFS observation, introducing
possible artifacts or errors in the derived quantities. Obviously, the S/N
threshold can be increased and as the cut is more restrictive these artifact
regions will diminish, but in some cases they will not completely
disappear. Furthermore, the increasing threshold may remove regions at the
boundaries of the object which are relevant for the scientific case.

The middle-right panel of \autoref{fig:vor} shows the selected spaxels after
applying the proposed S/N optimization method using the same central
wavelength and width as before, in this case the selected
regions follow very closely the morphology of the galaxy as shown in the
V-band continuum image, the boundaries are well defined and no additional
artifact spaxels are found in the FoV of the observation, as in the simple S/N
threshold case. The spectroscopic properties of this galaxy can be better
recovered from both the 1-D integrated spectrum and/or the 2-D spaxels of the
selected regions using the optimization technique. 
Note that the S/N optimization method works on spaxel-by-spaxel basis, it does
not intend to change the original spatial resolution of the input data, but
simply to characterize those regions in which relevant information can be
extracted. In that regard, the idea behind of the S/N optimization is
different to that of spatial adaptive binning techniques (such as the Voronoi
tessellation or adaptive smoothing) in which the final resolution of the
extracted spectra changes.

\begin{table*}
\caption{General science applications for different S/N optimization techniques within IFS.
  \label{tab1}}
\centering
\begin{tabular}{ll}
\hline\hline\\[-5pt]
S/N technique & General science applications within IFS\\
\hline\\[-3pt]
S/N optimization (based on the emission lines)      & Emission lines in poorly resolved high-z SFGs, objects with weak nebular emission,\\[2pt]
                                                    & intensity and kinematics of emission line ionized gas\\[2pt]

S/N optimization (based on the continuum)           & Absorption feature properties, SSP fitting, age-metallicity indices\\[2pt]

S/N optimization (removing systemic velocity field) & Average local kinematic properties of objects with evidence of shocks and turbulence\\[2pt]

Voronoi tessellation                                & Stellar kinematics, stellar absorption line strengths, dynamical modelling, \\[2pt]
                                                    & velocity fields of objects with a rich structure in their innermost (brightest) regions \\[2pt]

S/N threshold cut                                   & Average properties of bright/faint regions\\[2pt]

\hline
\end{tabular}
\end{table*}

Nevertheless, more sophisticated techniques that imply a spatial adaptive
binning rely on a proper calculation of the quantity that represents the
intensity of a signal and its ratio with the corresponding noise. An
inaccurate input proxy would result in a incorrect binning and a subsequent
erroneous analysis of the recorded signal. This is the case of the Voronoi
tessellation binning \citep{Cappellari:2003p4046}, a popular tool that has
found many applications in astrophysics \citep{Cappellari:2009p3997}. The
bottom-panels of \autoref{fig:vor} show the Voronoi binning segmentation of
the selected spaxels using the S/N threshold technique (left) and the S/N
optimization (right), in both cases the target minimum S/N within each bin was
5. In both cases, the Voronoi technique segments small and similar bins in the
the central spaxels of the object, but in the outer regions the bins become
much larger and not spatially related in the S/N threshold case compared to
the S/N optimized bins, which are much compact and smaller, fact that is
evident just from a visual inspection. In summary, the application of the S/N
optimization method can be very advantageous compared to a simple S/N
threshold cut, also the technique can be implemented to determine a suitable
input proxy for more sophisticated adaptive binning techniques, such as the
Voronoi tessellation.

\section{Conclusions}
\label{sec:4}

In this paper we propose a method to extract integrated spectra from a 2D IFS
data cube based on a S/N optimization by employing a procedure relying on
either continuum (stellar) or line (nebular) emission features from the
final-reduced observed spectra. The methodology of the spectra selection is
based on a {\em statistical} S/N defined in a different manner for the
continuum and line-emission cases. The actual implementation of the extraction
method relies on stacking the S/N-sorted spectra and calculating the 
{\em cumulative} S/N of the integrated spectrum. 
We have shown several examples to illustrate qualitatively and quantitatively
the potential of this optimization method for studies based on both,
continuum and emission features. As for the nebular emission, we have shown
a significant better definition of the kinematic profiles for the optimized
spectrum. Regarding the weak stellar absorption features, we have found a
significant better constrain of the output solution when applying the SSP
fitting technique. In particular, Monte Carlo simulations in a particular
case studied show that the uncertainties in the output Z, and $A_V$ are
reduced by factors of 5 and 2, respectively, when compared with the
unprocessed case.

We provide series of IDL scripts that can be applied to basically
any IFS datacube in order to implement the S/N optimization here
described. This tools are made available to the community as part of the
{\sc PINGSoft} IFS software package.
In addition to the proposed method to optimize the S/N of the integrated
spectrum, more standard procedures to obtain integrated spectra are possible
with mentioned tools. In particular the following options are
implemented:

\begin{enumerate}
\item[i)] Integration of the spectra over a specific geometrical region of the galaxy/FoV
\item[ii)] Integration of the spectra above a given S/N threshold
\item[iii)] A combination of i) and ii), or
\item[iv)] Integration of the spectra that optimize the S/N over a given
  region, following the method here described.
\end{enumerate}

\noindent
These possibilities can be done on the basis of a (S/N)$_c$ or a (S/N)$_{em}$, with
and without the velocity field correction (see Appen. \ref{appen:1}).

The process for obtaining an integrated spectrum for an optimal science use is
complex and it requires in practice a trade-off among several variables
(e.g. S/N, geometrical region probed, spatial resolution, etc). 
Interactive user-friendly versatile tools, as the one presented here, are
essential to carry out this process in an efficient manner.
The application of adapting techniques in order to extract integrated
information in future 2D spectroscopic surveys 
might be relevant when dealing with objects that are in a photon starving
regime, as it is usually in high-z studies, but it is also relevant when
analysing faint features in local bright objects, and in cases where the
analysis is very sensitive to S/N.
\autoref{tab1} summarizes some general science scenarios in which the S/N
optimization technique can be applied, together with the current main
applications of other techniques. The S/N optimization based on emission lines
could be advantageous when studying emission lines in poorly resolved high-z
star forming galaxies or objects with weak emission lines, and when the
systemic velocity field is removed, the method can be applied to objects with
evidence of shocks, turbulence, AGNs, etc. to decouple the kinematic
components of the integrated spectrum of the galaxy. The S/N optimization
based on the continuum may help to retrieve weak continuum features in the
integrated spectra of galaxies allowing to constrain in a more efficient way
the derived physical parameters of the unresolved stellar populations using
techniques such as spectral synthesis or age-metallicity sensitive absorption
features. In all cases, the S/N optimization allows to select a subsample of
spaxels to perform a spatially resolved studies and/or to use the outputs as a
proxy for more sophisticated adaptive binning techniques.
Note that the S/N optimization here proposed might be also applicable to IFS
observations of nearby objects or any other 2D observation where S/N-sorted
spectra may be relevant. The ultimate application of this method depends on
the nature of the IFS data, the scientific case and/or the practical needs of
a potential user.


\begin{acknowledgements}
  We would like to acknowledge S. F. S{\'a}nchez for his valuable comments and
  support regarding the SSP fitting techniques and the analysis of the Monte
  Carlo simulations.
  This work has been supported by the Spanish Ministry of Science
  and Innovation (MICINN) under grant ESP2007-65475-C02-01.
  FFRO acknowledges the Mexican National Council for Science and Technology
  (CONACYT) for financial support under the programme Estancias Posdoctorales
  y Sab{\'a}ticas al Extranjero para la Consolidaci{\'o}n de Grupos de
  Investigaci{\'o}n, 2010-2011.
  We would like to thank the anonymous referee for the very valuable comments
  and suggestions which improved the final content of this paper.
\end{acknowledgements}

\bibliographystyle{aa}
\bibliography{aa}


\begin{appendix}
\section{S/N optimization scripts: working example}
\label{appen:1}

In this section we present a working example of the implementation of the S/N
optimization described in this paper using the {\tt vfield\_3D.pro} and 
{\tt s2n\_optimize.pro} routines of the IDL {\sc PINGSoft} library.
Let us take as an example the VLT-VIMOS IFS observation of the LIRG IRAS
F06295-1735 \citepalias{Arribas:2008p3550} as shown in \autoref{fig:vimos}. 
The strength of its intrinsic velocity field can be determined by examining
the (observed) \ha\ emission line wavelength peak with respect to
reference (nuclear) spectrum. In the case of IRAS F06295-1735, we find that
the maximum displacement of the peak of \ha\ is about $\pm 5$ \AA, which
translates to $\approx 230$ km\,s$^{-1}$, which is more than a factor of two
larger than the effective spectral resolution for this particular observation
\citepalias[$1.83 \pm 0.25$ \AA\ (FWHM), $87 \pm 12$ km\,s$^{-1}$ at $\lambda$6300,
see][]{Arribas:2008p3550}. Therefore, a velocity-field correction is
recommended.

Assuming a working installation of IDL + {\sc PINGSoft}, a flux and wavelength
calibrated 3D cube FITS file of the target {\tt IRAS06295.fits}, and using as
a reference a nuclear spectrum with ID 770, on which the \ha\ emission
line is centered at the observed (redshifted) wavelength of 6700 \AA, the
velocity field correction is performed by applying the following command:

{\scriptsize
\begin{verbatim}
IDL> vfield_3D, 'IRAS06295.fits', 770, 6700 
\end{verbatim}
}

\noindent
The routine performs a wavelength shift by applying a cross-correlation with
respect to the reference spectrum, saves the distribution of the wavelength
shifts across the FoV of 3D cube and shows a histogram of this 
distribution. The script outputs the file: 
{\tt IRAS06295.vfield.fits}. Additional options that control minor details in
the operation are also available, see Appen. \ref{appen:2}.

Once the target has been corrected for its instrinsic velocity field we can
proceed to extract spectra using the S/N optimization described in this
paper. Let us assume that we want to extract the spectra based on the S/N on
the continuum using a narrow band centered at 6200\,\AA\, and in the case of
the S/N on emission, we want to base the calculation on the \ha\ emission
line region centered at 6700\,\AA\ (observed). Using this information, we can
simply proceed to implement the S/N optimization by applying the following
command:

{\scriptsize
\begin{verbatim}
IDL> s2n_optimize, 'IRAS06295.vfield.fits', 6200, 6700 
\end{verbatim}
}

\noindent
The routine will display on the screen a window showing the relevant spectral
regions for both the continuum and emission region, emphasising the narrow
bands actually used to obtain the S/N on both cases. The width and position of
the continuum and emission line pseudo-continuum adjacent bands can be
controled by using the {\tt WIDTH}, {\tt WCONT} and {\tt PSEUDO} options. If
the shown spectral regions satisfy the user, the routine proceeds to calculate
the S/N on the continuum and emission line feature, and shows a window with
two rows, on the top three plots corresponding to the continuum case, and on
the bottom to the emission line case. The left panels shows the spectra sorted
in decresing order of S/N, as shown in \autoref{fig:snr}, the vertical line
corresponds to the threshold cut value (S/N $> 1$); the middle panels show
the cumulative S/N on the integrated spectra, as shown in \autoref{fig:integ},
the vertical line shows the threshold value where the maximum S/N is found for
each case. The regions corresponding to the left of this threshold are shown
on the right panels, showing a visualization of the input 3D cube as a
narrow-band image with a certain intensity scale, as shown in
\autoref{fig:vimos}.

At this point, the user can either accept or modify interactively either the
threshold values for each sample 
and/or the type of continuum noise calculation ($N$ vs. $\sqrt{N}$). 
The selected regions will be updated after
each iteration, so the user has full visual control of the final selected
regions. Once the threshold values are set, the script extracts the selected
spectra and writes a Row-Stacked-Spectra FITS file and corresponding position
table for both the continuum and emission line samples. The routine produces a
2D map with of the S/N values calculated for each spaxel on
both samples and some additional output files (shown in the terminal window)
that can be used as input proxies for more sophisticated adaptive binning
techniques (see \autoref{sec:vor}).

Additionally, the {\tt s2n\_ratio\_3D} code extracts spectra interactively based on a
continuum and/or emission line S/N floor, the syntax and input parameters are
exactly equal as for the {\tt s2n\_optimize} command. The user can omit the
interactive mode by defining the {\tt S2N\_MIN} parameter as a two-entries vector
containing the S/N threshold for the continuum and emission line samples 
{\tt [S2N\_c,S2N\_e]}, i.e.

{\scriptsize
\begin{verbatim}
IDL> s2n_ratio_3D, 'IRAS06295.vfield.fits', 6200, 6700, S2N_MIN=[20,100]
\end{verbatim}
}

\noindent
This command line will extract all spaxels with (S/N)$_C$ $\ge 20$ and
(S/N)$_{e}$ $\ge 100$. For a description of all available options see
Appen. \ref{appen:2} or the {\sc PINGsoft} user guide.

\end{appendix}

\begin{appendix}
\section{Routines syntax}
\label{appen:2}

\noindent
Calling sequence for the {\tt vfield\_3D.pro} routine:

{\scriptsize
\begin{verbatim}
vfield_3D, 'OBJECT.fits', ID_reference, eline_reference [, out_str, 
            CONT1=cont1, CONT2=cont2, EXTENSION=extension, PT=pt, 
            PREFIX=prefix, LMIN=lmin, LMAX=lmax, /PS, /FORCE_FIT ]


  'OBJECT.fits':  String of the wavelength calibrated RSS 
                  or 3Dcube FITS file

   ID_reference:  ID of the reference spectrum in the mosaic 
                  (as seen by VIEW_3D)

eline_reference:  Wavelength of the emission line to use in order 
                  to perform  the cross-correlation
                  e.g. Halpha 6563*(1+z)

OPTIONAL

  out_str: Name of the optional output structure (including .vfield,
           ie. the shift applied to the spectra with respect
           to the reference spectrum)

EXTENSION: Non-negative scalar integer specifying the FITS extension 
           to read

       PT: String of the position table of the RSS file in ASCII 
           format (compulsory if the RSS+PTable names are not in 
           the standard format and/or they are not included in the 
           default instruments/setups)

   PREFIX: String with the prefix of the output files (if not set, 
           the program constructs a string based on the input 
           information)

     VMAX: Maximum velocity (km/s) with respect to the reference 
           spectrum for which a wavelength shift is applied. Spectra 
           with larger values are not shifted. Default: 300 km/s

  CONT1/2: Continuum bands used to normalize the spectra, 
           default: eline +/- 80 Ang

 LMIN/MAX: min/max wavelength for the cross-correlation, 
           default: eline +/- 100 Ang

      /PS: Writes a Postscript file of the wavelength shift 
           histrogram

/FORCE_FIT: Forces a Gaussian fit to each spectra before  
            calculating the cross-correlation (NOT SUGGESTED)
\end{verbatim}
}
\vspace{0.5cm}

\noindent
Calling sequence for the {\tt s2n\_optimize.pro} routine:

{\scriptsize
\begin{verbatim}
s2n_optimize, 'OBJECT.fits', lam_cont, lam_eline  [, out_str,   
               WIDTH=width, WCONT=wcont, PT=pt, EXTENSION=extension,
               PSEUDO=[lam_pseudo1,lam_pseudo2], /SQRT, 
               PREFIX=prefix, FILTER=filter, BAND=band, CT=ct, 
               FONT=font, /SILENT, /DRAW, /LINEAR, /GAMMA, 
               /LOG, /ASINH ]

'OBJECT.fits':  String of the wavelength calibrated RSS 
                or 3Dcube FITS file.

     lam_cont:  Central (observed) wavelength of the (featureless) 
                continuum band to calculate the (S/N)_cont

    lam_eline:  Central (observed) wavelength of the emission line 
                feature band to calculate the (S/N)_eline

OPTIONAL

    out_str:  Name of the optional output structure.
 
      WIDTH:  Width (in Angstroms) of the spectral regions 
              (elines and pseudo-continuum bands) from where the 
              (S/N)_eline is obtained. Default: 50

      WCONT:  Width (in Angstroms) of the continuum band from  
              where the (S/N)_cont is obtained. Default: 50

     PSEUDO:  A two-entries vector [lam_pseudo1,lam_pseudo2] 
              containing the central wavelenghts of the 
              pseudo-continuum adjacent bands used to calculate 
              (S/N)_eline. Default: [lam_eline-100,lam_eline+100]

     /SQRTN:  When this keyword is set, the continuum noise is  
              equal to sqrt(sigma). This may be useful when the
              noise is strongly non Poissonian, if the structure 
              of the signal in the spaxels is not optimally  
              weighted and/or if there are strong gradients in 
              the S/N.

  EXTENSION:  Non-negative scalar integer specifying the FITS  
              extension to read.

         PT:  String of the position table of the RSS file in 
              ASCII format (compulsory if the RSS+PTable names 
              are not in the standard format and/or they are not
              included in the default instruments/setups)

     PREFIX:  String with the prefix of the output files. 
              (if not set, the program constructs a string based
              on the input information)

    /SILENT:  Skips wavelength range confirmation and prints 
              minimum output information on-screen

Visualization options:  

  FILTER, BAND, CT, FONT, /DRAW, /LINEAR, /GAMMA, /LOG, /ASINH 

as explained in VIEW_3D.pro
\end{verbatim}
}
\vspace{0.5cm}

\noindent
Calling sequence for the {\tt s2n\_ratio\_3D.pro} routine:

{\scriptsize
\begin{verbatim}
s2n_ratio_3d, 'OBJECT.fits', lam_cont, lam_eline [, out_str, 
               S2N_MIN=[s2n_cont,s2n_eline], WIDTH=width, 
               WCONT=wcont, PSEUDO=[lam_pseudo1,lam_pseudo2], 
               /SQRTN, PT=pt, PREFIX=prefix, EXTENSION=extension, 
               FILTER=filter, BAND=band, CT=ct, FONT=font, 
               /SILENT, /DRAW, /LINEAR, /GAMMA, /LOG, /ASINH ]

'OBJECT.fits':  String of the wavelength calibrated RSS 
                or 3Dcube FITS file

     lam_cont:  Central (observed) wavelength of the (featureless) 
                continuum band to calculate the (S/N)_cont

    lam_eline:  Central (observed) wavelength of the emission line 
                feature band to calculate the (S/N)_eline

OPTIONAL

      S2N_MIN:  A two-entries vector [s2n_cont,s2n_eline] containing 
                the S/N threshold values.

              NOTE: when this parameter is included, the script 
                    extracts automatically those spaxels above the 
                    input values for both samples without 
                    interaction.

Similar input parameters as in S2N_OPTIMIZE.pro
\end{verbatim}
}

\end{appendix}

\end{document}